# Air-Water Interface-Assisted Synthesis and Charge Transport Characterization of Quasi-2D Polyacetylene Films with Enhanced Electron Mobility via Ring-Opening Polymerization of Pyrrole

Kejun Liu[1], Nadiia Pastukhova[2], Egon Pavlica[2], Gvido Bratina[2] et al., Xinliang Feng[1]

[1] Faculty of Chemistry and Food Chemistry & Center for Advancing Electronics Dresden, Technische Universität Dresden, 01062 Dresden (Germany)

[2] Laboratory of Organic Matter Physics, University of Nova Gorica, Vipavska 13, Nova Gorica SI-5000, Slovenia

## Abstract

Water surfaces catalyze some organic reactions more effectively, making them unique for 2D organic material synthesis. This report introduces a new synthesis method via surfactant-monolayer-assisted interfacial synthesis on water surfaces for ring-opening polymerization of pyrrole, producing distinct polypyrrole derivatives with polyacetylene backbones and ionic substitutions. The synthesis result in quasi 2D polyacetylene (q2DPA) film with enhanced charge transport behavior. We employed time-of-flight photoconductivity (TOFP) measurements using pulsed laser light of tunable wavelength for photoexcitation of the charge carriers within the q2DPA film. The charge transport was measured in the lateral direction as a function of external bias voltage ranging from 0 V to 200 V. We observed high electron mobility (μ) of q2DPA reaching values of 375 $cm^2$ $V^{-1}$ $s^{-1}$ at bias voltage $V_b$ = -20V and photon energy of 3.8 eV.

# Introduction

Water is the most crucial liquid in nature, and research has shown that water surfaces uniquely catalyze various organic reactions, such as Diels-Alder and Claisen rearrangement reactions, far more effectively than common organic solvents.[1–3] This acceleration is attributed to surface hydrogen bonding effects, highlighting the distinctiveness of the water surface as a reaction environment compared to bulk water.[4] Following this understanding, the exploitation of water surface chemistry for the synthesis of new materials has increasingly garnered attention in recent years. Moreover, materials synthesized on the water surface exhibit advantages such as large-area uniformity and ease of transfer, making them well-suited for practical applications. These characteristics are pivotal for the scalable production and integration of these materials into a broad spectrum of devices and systems, further enhancing their applicability in real-world scenarios.

In this report, we describe a new method based on the ring-opening polymerization of pyrrole by surfactant-monolayer-assisted interfacial synthesis (SMAIS) on water surface. In bulk solution, the polymerization of pyrrole to produce polypyrrole, one of the oldest conducting polymers, has been well-established.[5] However, we found that polymerization on water surface yields distinct products. The synthesis involves the open-ring procedure, resulting in conjugated polymer formation with a backbone of polyacetylene and substitution of -OH and -$NH_3^+$ groups. Furthermore, a surfactant monolayer can guide the polymerization to yield large-area crystalline q2DPA films on water surface by SMAIS method.[6] The q2DPA film contains a PA backbone of fully stretched conformation via ionic interchain linkers, thus favoring the coplanarity of the sp2-hybridized carbon rings. The q2DPA preserves its semiconducting properties even though it is doped by $SO_4^{2-}$. Moreover, q2DPA shows rapid photoconductivity response and record-high charge carrier mobility 375 cm$^2$ V$^{-1}$ s$^{-1}$ measured by the in-plane time-of-flight photoconductivity method.

## Results and discussion

**On-water surface ring-opening reaction**

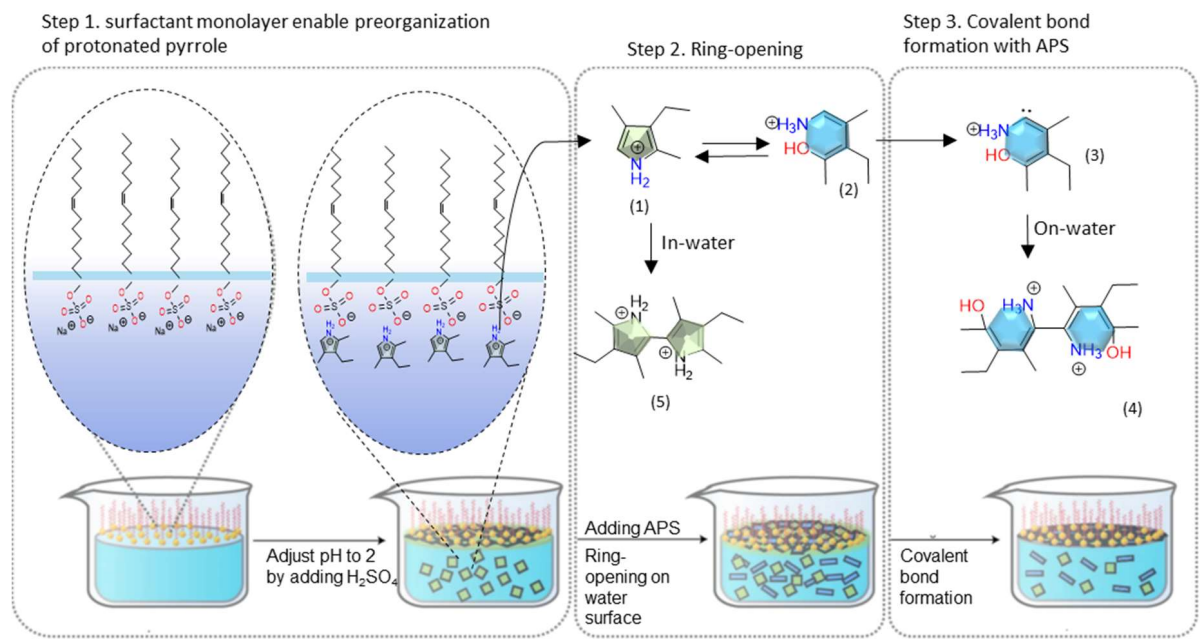

***Figure 1.*** *Schematical illustration of the steps 1–3 of model reactions of 3-ethyl-2,4-dimethylpyrrole (1) to form a diammonium product (4) by the SMAIS method using ammonium persulfate as an oxidizing agent (1 equivalent to M1) in an acidic condition ($H_2SO_4$, pH = 2).*

To examine the feasibility of this ring-opening reaction for on-water surface polymerization, we evaluated a model reaction of 3-ethyl-2,4-dimethylpyrrole (**1**) to form a diammonium product (**4**) by the SMAIS method using ammonium persulfate as an oxidizing agent (1 equivalent to M1) in an acidic condition ($H_2SO_4$, pH = 2), as shown in Figure 1. The reaction occurred in three steps; first, the preorganization of compound **1** guided by the surfactant monolayer, second, Lewis acid-catalyzed reversible ring-opening and third, covalent bond formation under oxidation. In step 1, a sodium oleyl sulfate (SOS) monolayer was prepared on a 25 mL aqueous solution of $H_2SO_4$ (9.33 mmol) in a beaker (diameter, 10 cm). The pH value of the subphase is 1.4. After 30 min, compound **1** (25 μmol) was slowly injected into the subphase, where it is protonated and dissolved well. And the pH of the subphase was changed to 2. As a consequence of the electrostatic interaction between the protonated compound 1 and the anionic head groups of SOS, compound **1** was readily adsorbed underneath the SOS monolayer. After 30 min, a 2 mL aqueous solution dissolving 11.41 mg ammonium persulfate (APS, 2 eq. to **1**) was injected into the aqueous subphase. Subsequently, there is a shining film forms on water surface after 12 h. The film can be fished out onto $SiO_2$(300 nm)/Si substrates for drying, and then dissolved in chloroform to further characterization.

The reaction mechanism of the model reaction is illustrated schematically in Figure 1. According to the scheme, compound **1** and its ring-opening product compound **2** can reach a balance in the aqueous solution. And compound **1** is the dominant species over compound **2** before the addition of APS. After the addition of APS, **2** can readily be converted to intermediate compound **3** and the final product compound **4**. The procedure efficiently moves the balance between **1** and **2** in the right direction to produce more **2**. Because **4** has strong interaction with the surfactant monolayer, **4** can be aggregated on the water surface to form a very obvious film under the guidance of the surfactant.

We performed electrospray ionization mass spectrometry (ESI-MS) to monitor the product evolution in steps 2 and 3. Both samples from the water subphase and film on the water surface were measured, which are denoted as in-water and on-water, respectively. MS spectrum presented peaks at maximum $m/z$ position are different in in-water reaction (Figure 2a) and on-water reaction (Figure 2b). The in-water spectrum shows a maximum $m/z = 717.22$, while the on-water spectrum presents a maximum $m/z = 495.35$. The peak at $m/z = 717.22$ corresponds to a product as shown in Figure 2c. It corresponds to a cluster containing two molecules of compound **5**, two sulphuric groups and two water molecules (theoretical $m/z = 718.92$). In contrast, the on-water reaction spectrum doesn't include this peak at the detectable range. Instead, the on-water sample has a maximum $m/z = 495.35$, revealing the cluster containing the ring-opening product (theoretical $m/z = 494.16$, Figure 2d). We also note that the peak of the in-water reaction with the strongest intensity has the $m/z = 140.01$, revealing the reaction from compound **2** to compound **3**. Although there is also compound **3** in the subphase, we cannot find the peak corresponding to compound **4** of the in-water sample in the detectable range of ESI-MS, which indicates that compound **4** can only exist on the water surface. Moreover, the electrospray ionization in ESI-MS measurement can break the products of the highest $m/z$ value to form fragments at the lower $m/z$ range, the peaks of which can also be assigned to fragments of expected products.

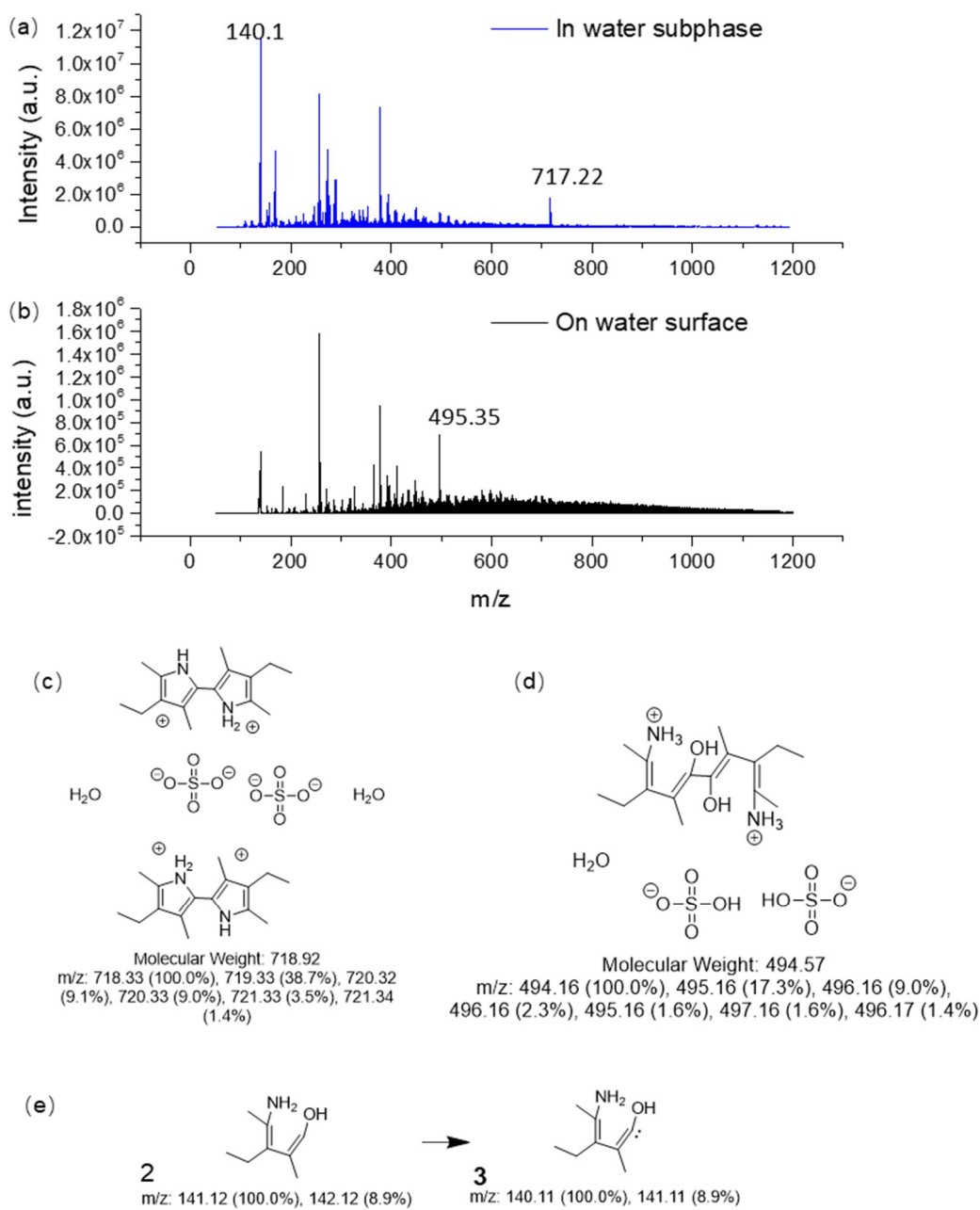

*Figure 3.* ESI MS spectra of (a) In-water and (b) On-water reactions. (c), (d) and (e)Chemical structures of detected cluster and the calculated m/z values.

To support the ESI-MS result, we performed matrix-assisted laser desorption/ionization–time-of-flight mass spectrometry (MALDI–TOF MS). However the aqueous solution cannot be used in MALDI–TOF MS measurement. So, we only measured the on-water sample. The product film of on-water was collected from the water surface and rinsed with water, dried at 80 °C, and then dissolved in chloroform to prepare the solution for the analysis by MALDI–TOF MS. After reaction for 12 h after addition of APS (step 3), the MS spectrum presented sharp peaks at $m/z$ = 397.4 (Figures 3a and b), which are assigned to a cluster containing compound **4**. The cluster has one molecule of **4** (Figure 3c), one sulphuric group, and one water molecule (calculated $m/z$ = 396.19).

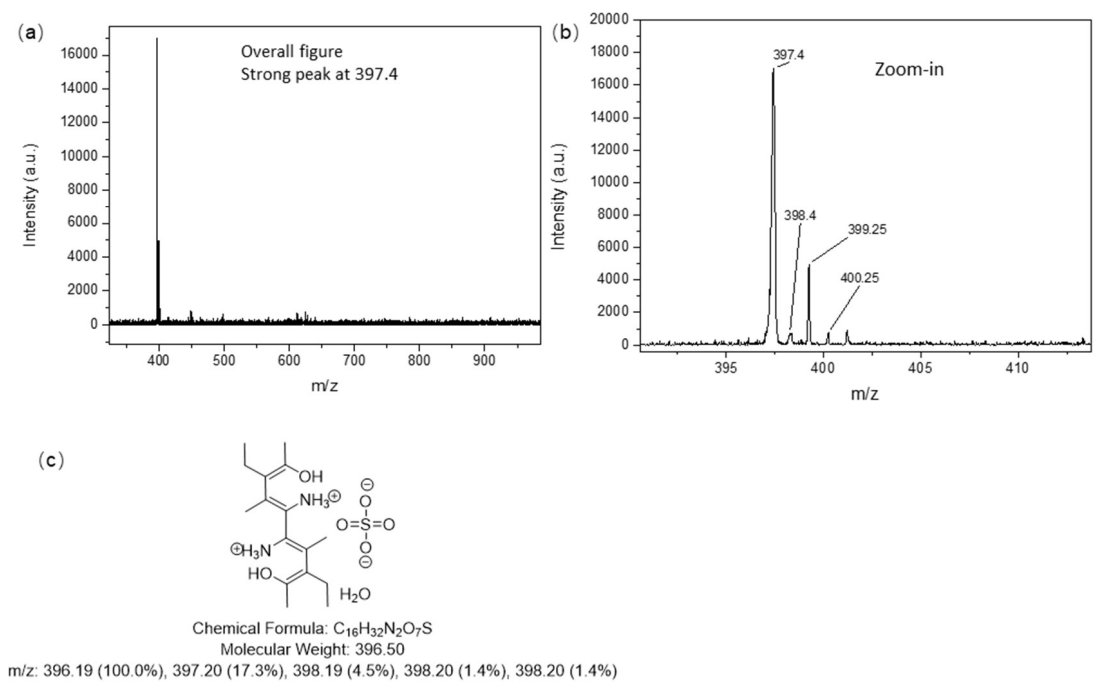

*Figure 3. (a) Overall (b) High-resolution MALDI-TOF MS spectra, (c) Chemical structure of detected cluster containing model compound 4.*

Furthermore, we measured FTIR and UV-vis spectroscopy with the on-water film to get more information about its chemical composition. As shown in Figure 4a, the -OH stretch of water was detected and presented as a strong and broad peak centered at 3440 cm$^{-1}$, alkyl chains of compound **4** were characterized as two peaks from the range of 2900 to 3000 cm$^{-1}$, corresponding to methyl and ethyl groups, respectively. At 1493 cm$^{-1}$, there is the peak reflecting C-H bending of conjugated structures. And S=O and S-OH stretch is also very intensive at about 1094 cm$^{-1}$. The UV-vis spectrum reveals two peaks at 281 nm and at 498 nm, which can be assigned to π → π* and n → π* transitions, respectively. It is similar to many aromatic molecules, indicating the aromatic nature of compound **4.**

In conclusion, ESI-MS, MALDI-TOF-MS, FTIR, and UV-vis experiments suggest that the 2D confinement at the air-water interface supported by the surfactant monolayers is imperative to enhance the reactivity to form ring-opening products on water surface.

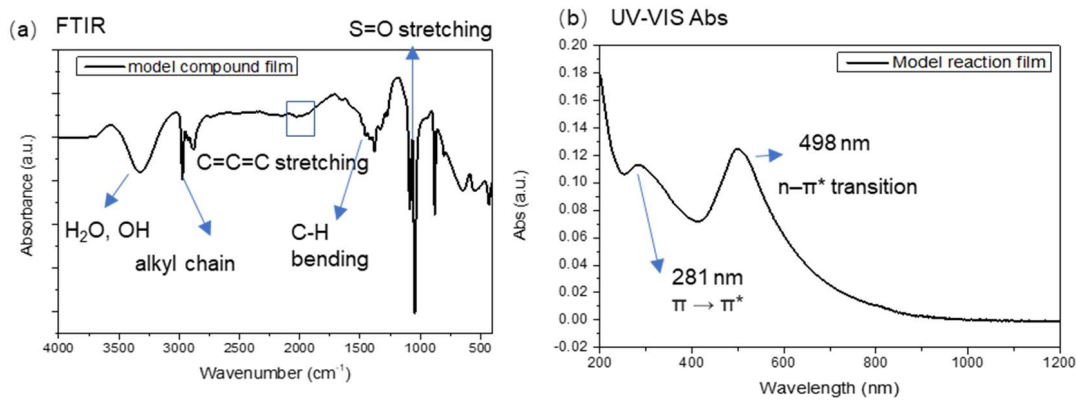

*Figure 4.* (a) FTIR, (b) UV-Vis of q2DPA film.

**Synthesis and characterization of quasi-2D polyacetylene**

Given the experimental observations and theoretical analyses of the open-ring reactions of the pyrrole ring, we then explored the synthesis of q2DPA from pyrrole by the SMAIS method as shown in Scheme 1. The reaction time was decreased to 2 h because pyrrole has higher reactivity than compound **1**. We used 0.5 equivalent sulphuric acid and 1 equivalent APS *versus* pyrrole. The accurate equivalent ratio is crucial to the synthesis of crystalline samples because sulphuric groups are needed to act as interchain linkers to build the quasi 2D lattice. Other procedures and conditions are the same as those of the model reaction. The expected product via the ring-opening polymerization is a quasi 2D polymer structure. Its backbone, as highlighted in Scheme 1, is a polyacetylene chain in a *cis* configuration. Individual polyacetylene chains are linked via sulphuric acid groups as interchain linkers.

According to the reaction mechanism revealed by the model reaction, the polymerization mechanism of q2DPA is illustrated in Scheme 2. Step 1 shows the reaction balance between pyrrole (compound **6**), 2H-pyrrol-1-ium (compound **7**) and (1Z,3Z)-4-hydroxybuta-1,3-dien-1-aminium (compound **8**). Compound **7** should be the dominant species in the subphase before polymerization. Compound **8** has a good affinity with surfactants because it has a cationic group and hydrogen bonds, which can be attracted by anionic surfactants (Step 2). And the absorption of surfactant can induce the balance moving to the right side to yield. After adding APS to induce the polymerization (step 3), compound **8** is consumed in the subphase to form films on water surface, which can move the balance to yield more ring-opening compound **8**.

It only took 30 mins to form a visible film. A longer reaction time (for example 2h) can induce the layer-stacking to form a thicker film (Step 4). Then, the film can be washed with chloroform to eliminate surfactant (Step 5), and dried for further measurement.

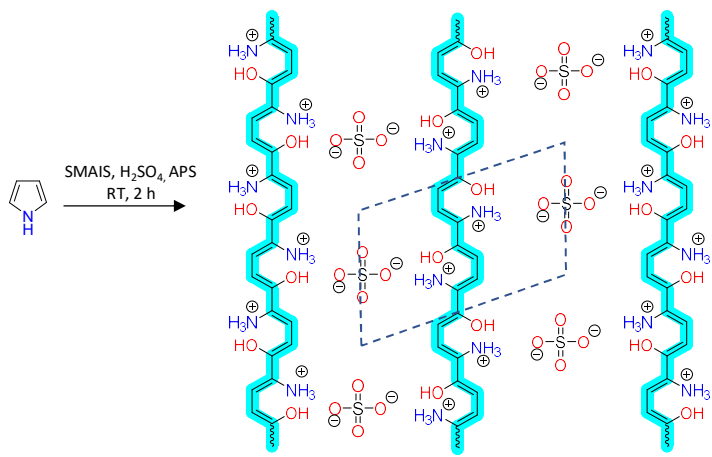

***Scheme 1.*** *Reaction scheme to synthesis q2DPA*

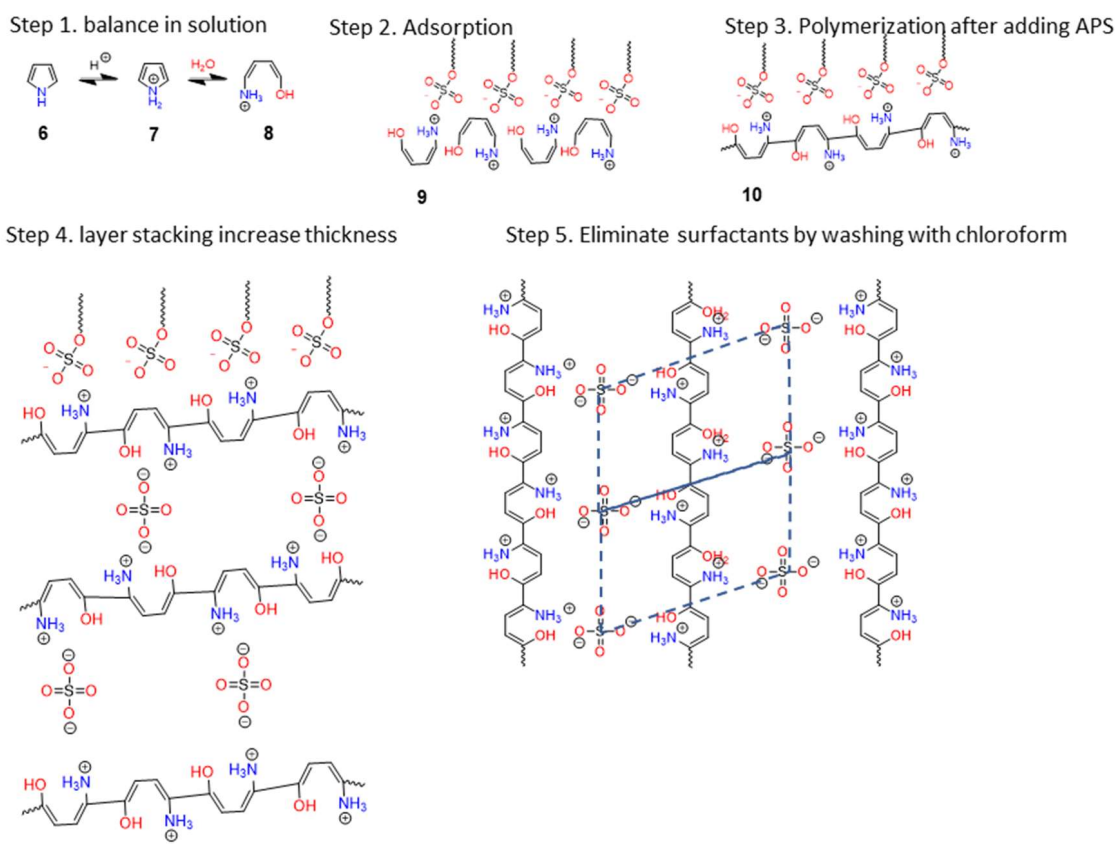

***Scheme 2.*** *Reaction mechanism scheme to synthesis q2DPA from pre-organized monomer to q2DPA.*

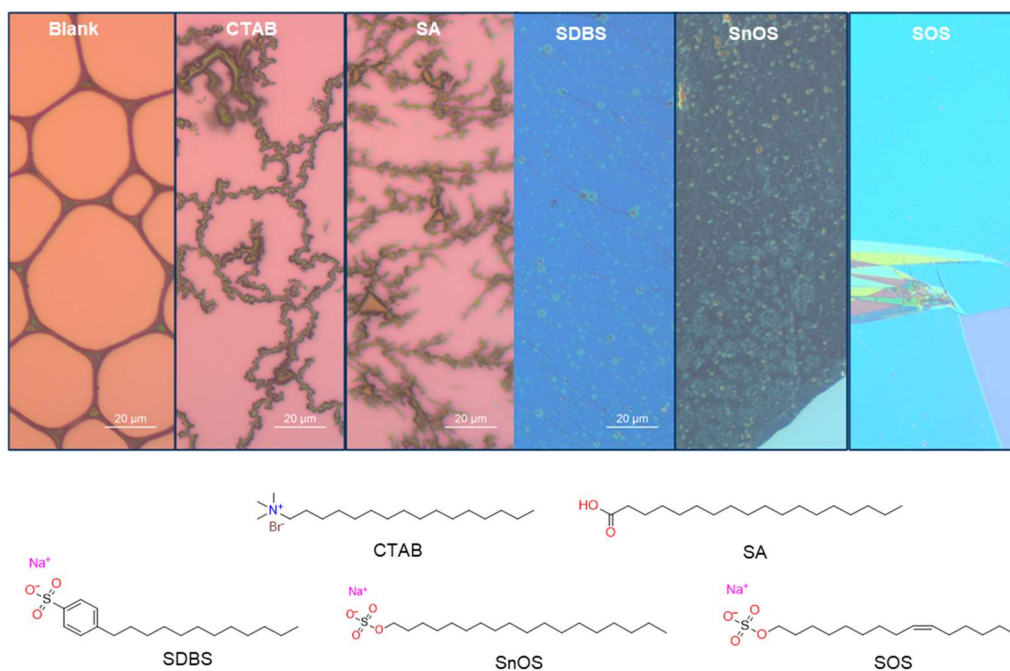

*Figure 5.* *Optical microscopic images of q2DPA films under different surfactants.*

According to the mechanism described above, the surfactant monolayer must play a crucial role in the synthesis of q2DPA film. To test the hypothesis, we used different surfactants, including CTAB, SA, SDBS, SnOS, and SOS for the SMAIS polymerization. For comparison, we also tried the synthesis without any surfactants, denoted as Blank, which cannot produce a continuous film. Cationic surfactant CTAB and neutral surfactant SA can yield fibrous morphology on the water surface. It indicates that the interaction between monomer and surfactants is electrostatic, not a hydrogen bond. Only anionic surfactants work well on inducing film formation.

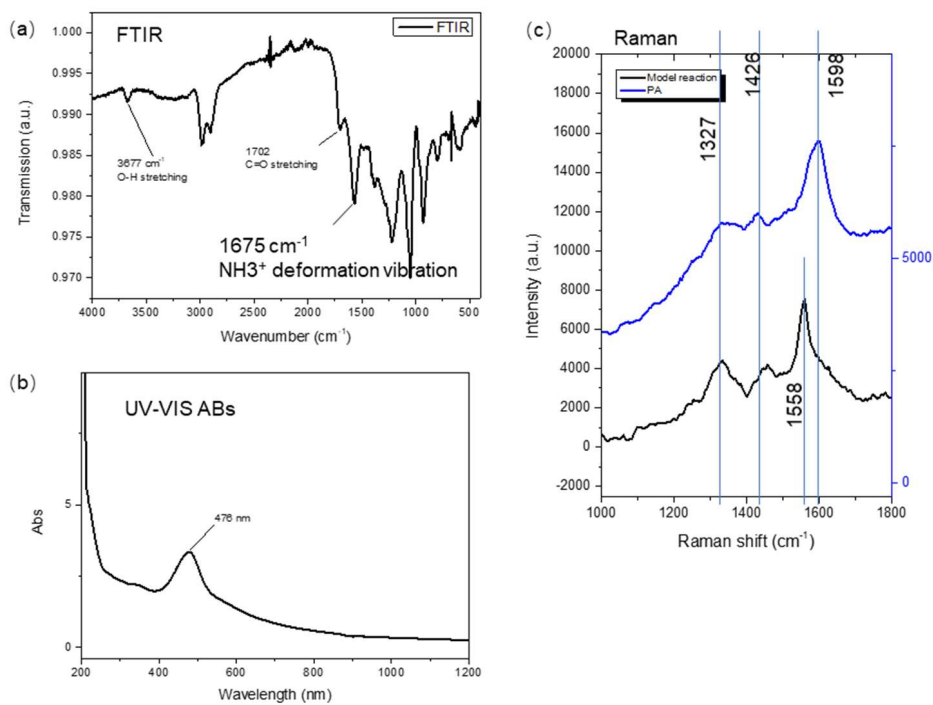

***Figure 6.*** *(a) FTIR, (b) UV-Vis, and (c) Raman spectra of q2DPA films.*

The FTIR spectrum (Figure 6a) of q2DPA presents a similar spectrum as that of the model compound, including the -OH stretch as a sharp band at a higher frequency at 3677 cm$^{-1}$. The high frequency of this vibration is characteristic of a strong OH bond with little or no hydrogen bonding, which can be contributed by the bonding of the sulphuric group and -OH. Besides, a peak at 1702 cm$^{-1}$ can be assigned to C=O, which can be caused by partial conversion from hydroxide groups to ketone groups. The sharp and strong peak at 1675 cm$^{-1}$ represents protonated amino groups. It should be noted that, unlike the -OH group, we cannot find the protonated amino groups were dynamically deprotonated because of the absence of a neutral amino band at about 3300 cm$^{-1}$. Compared with model compound **4**, the polymer shows a stronger n → π* transition peak at 478 nm with a blue shift of 20 nm. A study on the explanation of the UV-vis spectrum is still needed. The Raman spectroscopy shows a more straightforward comparison between q2DPA and model compound **4**. Both of them have a similar pattern, saying three peaks in the range from 1000 cm$^{-1}$ to 1800 cm$^{-1}$. The spectrum of q2DPPA matches the reports of doped PA. The C=C vibration of q2DPPA was found at 1598 cm$^{-1}$ (q2DPPA), which is higher than 1558 cm$^{-1}$ of model compound **4** because of higher conjugation in the polymer sample. A weaker peak at 1426 cm$^{-1}$ is representative of out-of-plane bands of C=C vibration. And 1327 cm$^{-1}$ reflects the stretching of hydrogen atoms in the structure.

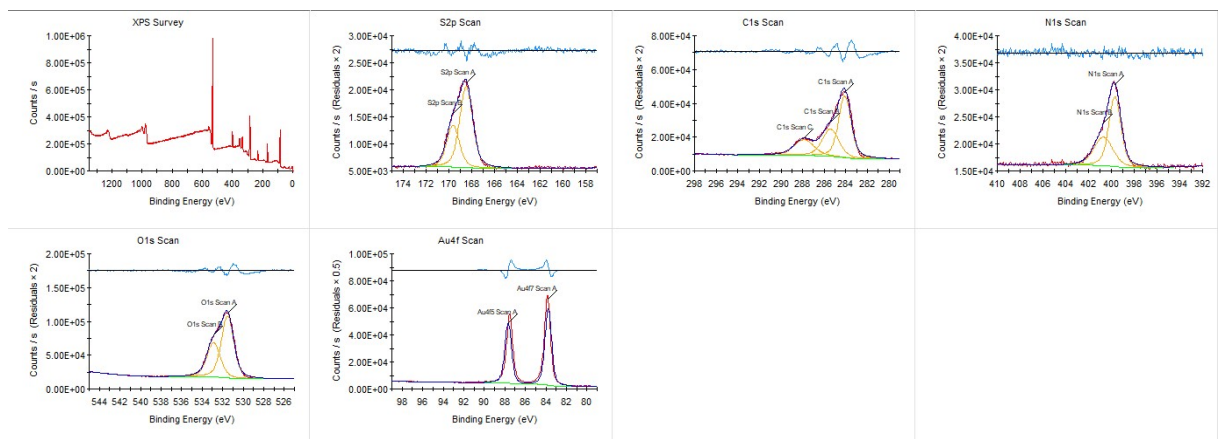

*Figure 7.* XPS spectra of q2DPPA, including XPS survey and high resolution spectra of S, C, N, and O species.

**Table 1.** Atomic ratio of different elements in XPS measurement.

| Peaks | Atomic % (%) |
|---|---|
| S2p Scan A | 4.9408 |
| S2p Scan B | 2.4754 |
| C1s Scan A | 23.345 |
| C1s Scan C | 9.5992 |
| N1s Scan A | 4.7119 |
| N1s Scan B | 3.0165 |
| O1s Scan A | 22.797 |
| O1s Scan B | 13.566 |
| Au4f7 Scan A | 1.7 |

To understand the chemical shift and composition of q2DPA, we used XPS spectroscopy. The survey spectrum proves that there are S, C, N, O, and Au species. Au is from the Au coating substrates, not the q2DPA film. S has two species, corresponding to $S2p_{1/2}$ (170.2 eV, Scan B) and $S2p_{3/2}$ (169.0 eV, Scan A) respectively. C1s can be divided into three subpeaks. Sp2 carbon shows the strongest peak at 284.0 eV (Scan A). The C-C component is set to the binding energy of 284.8eV (Scan B), which may be a signal from ubiquitous carbon contamination on samples forming during air exposure, so-called adventitious carbon (AdC). And 288.2 eV (Scan C) represents the band of C=O or C-OH. Regarding the N1s scan, The N1s peak with a maximum at about 399 eV (Scan A) suggests the existence of different C–N. Positively-charged nitrogen is obtained at B.E. = ~401 eV (N1s Scan B). O1s are also divided to show two peaks at 531.8 eV and 532.5 eV, which are assigned to the peak centers of C=O and C-O-H. Those XPS spectra show an atomic ratio of C/N/O/S = 32.93/7.72/36.35/7.43 (Table 1), which is close to the theoretical value of 4/1/5/1, according to the proposed chemical structure in Scheme 1.

The crystallinity of q2DPA was checked by pXRD with a film of 500 nm thickness. It proves the high crystallinity by its sharp peaks, at 12.5°, 24.0°, 36.0°, and 47.5°. Those peaks correspond to d space of 0.74 nm, 0.37 nm, 0.25 nm, and 0.22 nm. The pXRD spectrum is very similar to that of crystalline fully carboxylated polyacetylene[7], which indicates a similar crystalline structure.

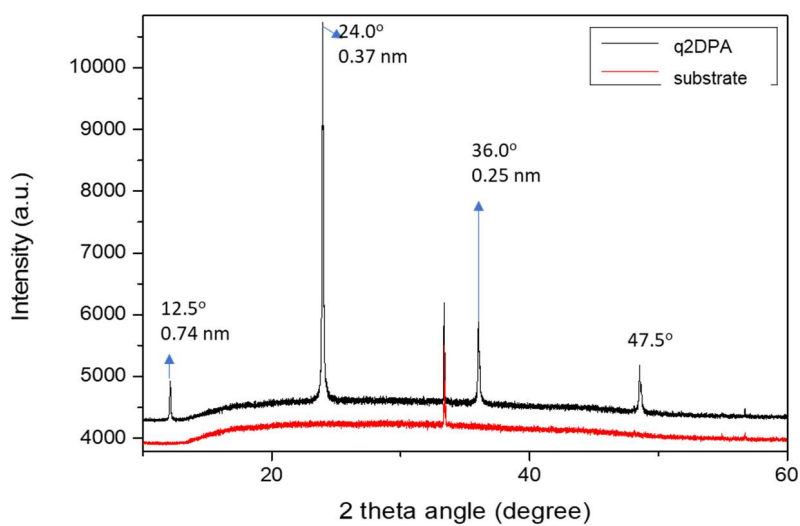

*Figure 7. pXRD pattern of q2DPA.*

TEM was also used to study the structure of q2DPA. However, q2DPA is too unstable under the beam due to its ionic structure. The large crystal domains were broken into tiny domains of 20 nm during the measurement. From the broken domains, we can still get the high-resolution image as shown in Figure 8. It shows a lattice fringe of 0.37 nm, corresponding to the most intensive peak of the pXRD pattern.

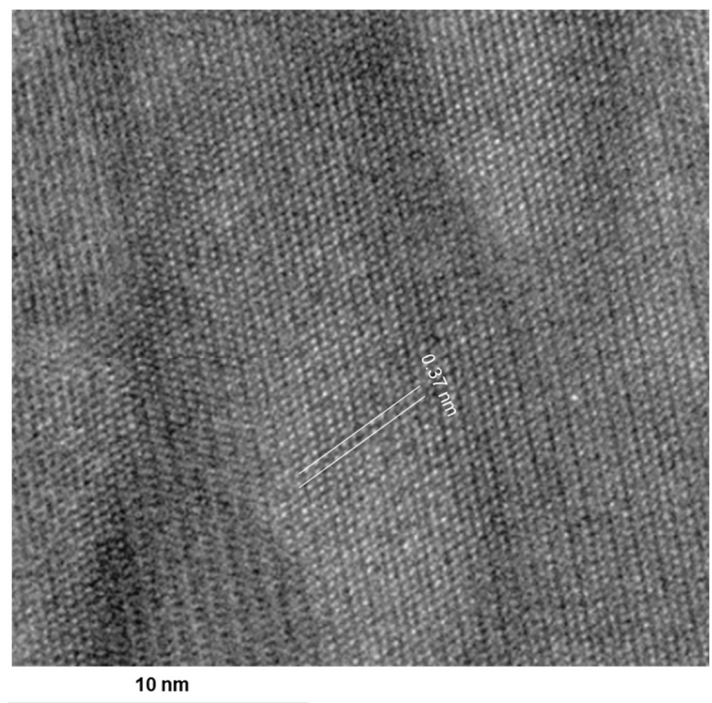

*Figure 8.* HR-TEM image of q2DPA, on the smaller domains of 20 nm, lattice fringe of 0.37 nm was observed.

**Charge transport study of quasi-2D polyacetylene films**

Observation of the high degree of 2D molecular arrangement of q2DPA films stimulated the study of charge transport through these films. Charge transport through films of π-conjugated polymers strongly depends on the polymer chain alignment and inter-chain arrangement[reference]. Linear, i.e. one dimensional (1D) conjugated polymers, such as poly(p-phenylene), poly(p-phenylene vinylene), poly(acetylene), polypyrrole, polythiophene, polyaniline, and their derivatives possess continuous π-electron delocalization along the chain axis.[8–12], resulting in high charge carrier mobility in that direction. Experimentally measured intrachain charge carrier mobilities exceed 100 cm$^2$ V$^{-1}$ s$^{-1}$ along the chain axis,[13] with reported values close to 600 cm$^2$ V$^{-1}$ s$^{-1}$.[14] Although the intrachain transport is reduced by chain structural disorder, monomer orientation disorder, and also chemical impurities, the critical bottleneck of the charge transport is caused by the low intrachain charge-transfer rates. Two-dimensional polymers extend π-conjugation to two dimensions, increasing thereby in-plane charge carrier mobility in these layers.[15,16] In contrast to one-dimensional polymers, the structure of q2DPA mimics the π-conjugation in the second dimension by sulphuric acid groups acting like interchain linkers (see Fig.1). And since the crystallinity of q2DPA layers was exceptionally high, as shown in Figure **Error! Bookmark not defined.**, we expected that

q2DPA thin-films exhibit a relatively low level of intrachain and interchain structural misalignment. Hence, interchain charge transport occurs through electronic states induced by interchain linkers. When different parts of the polymer layer spacially overlap which is also referred to as overlap integrals, it influences interactions or correlations between segments of the polymer layer. Higher overlap integrals generally lead to higher charge carrier mobilities in organic semiconductors. This is due to the strong electronic coupling facilitating more efficient delocalization of charge carriers, reducing localization effects that can impede mobility.

To study the charge transport in q2DPA we used the time-of-flight photoconductivity (TOFP) technique[17], to obtain a deeper understanding of the effect of these linkers on the macroscopic scale. TOFP was selected due to its versatility and because it is suitable for the study of charge carrier transport in the in-plane direction of thin layers. The details of TOFP were presented elsewhere.[17] Briefly, TOFP measurement is performed by exciting the charge carriers with a short light pulse in a layer of q2DPA. Photo-excited charge carriers are separated and collected by an external bias applied between two coplanar electrodes evaporated onto the q2DPA. For that purpose, q2DPA was placed on a quartz substrate and the two top-contact coplanar electrodes were deposited on top of the layer (Figure 9a).

Aluminum was found to form a blocking electrode with q2DPA and was used to avoid charge injection from the metal. The aluminum electrodes were deposited by thermal vacuum evaporation of Al through a shadow mask. The distance between the electrodes was 250 μm. From TOFP results (Figure 9c, d) it appears that mobility has an inverse dependence on the electric field, which is not typical for most OS. We used a 3 nanoseconds-pulse laser of a repetition rate of 3 Hz to generate electron-hole pairs in a thin film of q2DPA as schematically presented in Fig. 9a. The excitation wavelength was 325 nm and the light was focused through a cylindrical convex lens into a line of a width of 16μm along the right electrode (Fig. 9b). The laser pulse energy was adjusted with commercially available neutral-density optical filters. An external voltage pulse of amplitude $V_b$ was applied to the left electrode (biased electrode) for the charge separation. After that, the photogenerated charge carriers were collected by the electrodes in the presence of the external electric field. The drift of photogenerated charge carriers induced displacement current - photocurrent transient $I(t)$ which was measured by a current amplifier and an oscilloscope. We assume the electric field between two coplanar electrodes is uniform and depends on the applied voltage ($V_b$) and the interelectrode distance $E = V_b / L$. In Figure 9c we present $I(t)$ measurement in a q2DPA film as a function of $V_b$ in a double logarithmic plot. $V_b$ was applied in a range from 0 V to 200 V with a step of 20 V.

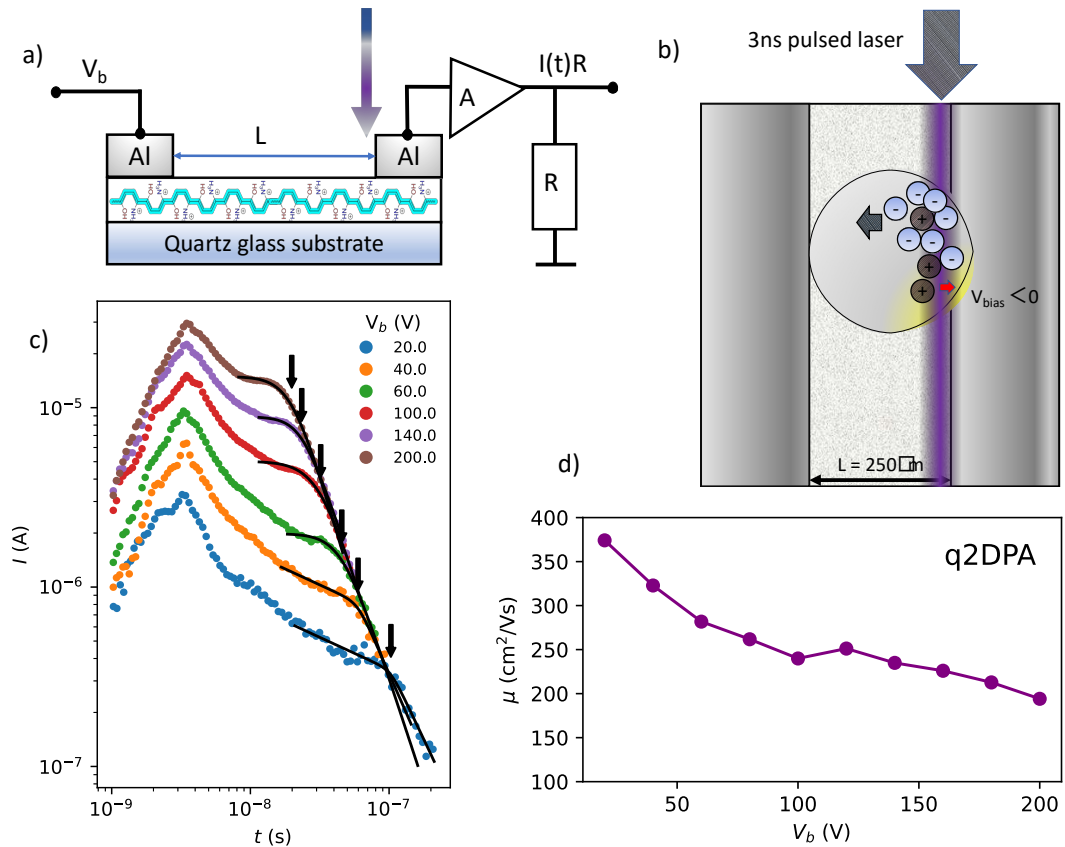

*Figure 9. (a) Schematic representation of a thin q2DPA layer and two coplanar Al electrodes on top of a quartz substrate. (b) Top-view schematic of two semi-infinite Al coplanar electrodes used for TOF measurement of q2DPA layer, purple line corresponds to the focused laser beam near one of the electrodes, blue arrows indicating the direction where the charge carriers are transported. (c) Time-of-flight photocurrent I(t) response to laser illumination of q2DPA as a function of a bias voltage $V_b$ in a double logarithmic plot, measured at 3.8 eV photon energy. (d) Mobility of q2DPA $\mu_{TOF}$ as a function of the $V_b$.*

The first peak of $I(t)$ at 3 ns in Figure 9c is not $V_b$-dependent and corresponds to the initial surge of photocurrent during laser illumination representing the photogeneration of free charge carriers and small polarons of both holes and electrons. After the first peak, $I(t)$ exhibit a non-monotonous decay in a double-logarithmic plot. A characteristic change of the $I(t)$ slope reflects the arrival of charge carriers towards the collecting electrode. From the polarity of $V_b$, we can determine, that positively charged holes were immediately collected to the illuminated right electrode, while electrons were collected across the q2DPA layer to the left electrode (Figure 9b). Compared to negative $V_b$, $I(t)$ exhibited monotonous decay without characteristic change of the slope and lower magnitude. Therefore, obtaining $I(t)$ only for positive $V_b$ demonstrates that q2DPA acts as an n-type material. The change of the $I(t)$ slope results from the arrival of the fastest charge carriers towards the left electrode and is referred to as the transit time ($t_{tr}$), which is used to calculate charge carrier mobility. $t_{tr}$ was determined using the $I(t)$ model (solid lines in Fig. 9c), which we introduced in Ref.[17] Our model empirically resolves the transition region in between two asymptotes, where $I(t)$ change its slope:

$$I(t) = I_0 (\frac{t}{t_{tr}})^{\alpha-1} \cdot (1 + (\frac{t}{t_{tr}})^{\gamma})^{\frac{-\beta}{\gamma}}$$

where $\alpha$ and $\beta$ determine $I(t)$ slope before and after $t_{tr}$. Both, $\alpha$ and $\beta$ are unitless parameters related to charge trapping and de-trapping dynamics. The value of $\alpha$ is restricted between 0 and 1 and was formally introduced as a dispersion parameter. While $\gamma$ represents the width of the transition region. It quantifies the distribution of transit times of the subgroup of those charge carriers, which induced the change of the $I(t)$ slope. From the measured transit times (arrows in Fig. 9c), we can obtain the effective mobility of the ensemble of the charge carriers:

$$\mu = \frac{\pi^2}{8} \frac{L^2}{V_b t_{tr}}$$

which stems from the assumption of charge carrier drifting with constant mobility in an electric field formed between coplanar electrodes when an external bias voltage is applied.

In Figure 9d we present the calculated mobility $\mu_{TOF}$ as a function of $V_b$. The latter is a zero-order approximate for the electric field. Notably, $\mu$ decreases with $V_b$. We can observe that the highest value of $\mu_{TOF}$ = 375 cm² V⁻¹ s⁻¹ is obtained at 20 V. With the increase of $V_b$ to 100 V, $\mu_{TOF}$ drops to 240 cm² V⁻¹ s⁻¹. The lowest value $\mu_{TOF}$ =190 cm² V⁻¹ s⁻¹ was measured at 200 V. An inverse relationship between mobility $\mu$ and the electric field is seldom observed, but it results as a characteristic fingerprint of the band-like transport in thin crystalline semiconducting films.[18,19] The structural disorder in imperfect thin-film crystals, introduced by defects or grain boundaries, can produce localized states and cause variations in the mobility edge. An additional important source of localized states in thin films can be the disorder due to impurities or surface roughness. Although their density is low, localized states can act as traps of charge carriers, reducing charge mobility. In thin films, charge transport is inherently percolative, with percolation paths rearranging from isotropic to predominantly directional with an increasing electric field.[19] This implies that effective trapping time depends on the electric field in a non-trivial manner. The effective trapping time increases with the electric field, presumably due to dead-end percolation routes, leading to an inverse relationship between mobility and the electric field. The above theoretical model fits our q2DPA films, which exhibit charge mobility in the range of typical band-like materials. In addition, pXRD (Figure 7) and HR-TEM (Figure 8) results show that q2DPA has a highly crystalline structure with the domain size of hundreds of micrometers. Hence, the inverse relationship between mobility and the electric field could be attributed to the rearranging of percolation paths with increasing electric field. Most importantly, the measured charge mobility ($\mu_{TOF}$) corresponds to the macroscopic transport in the range of 250 μm.

## Conclusion

In this manuscript, we present recent progress in the synthesis of semiconducting quasi-2D polyacetylene film of record-high mobility via ring-opening polymerization of pyrrole by the SMAIS method on the water surface. Firstly, we found a new on-water-surface reaction that supports the open-ring polymerization of pyrrole to build a polyacetylene backbone from pyrrole. Then, the analysis of chemical composition and crystal structure is presented to support the q2DPA. The q2DPA film shows record-high electron mobility of 375 $cm^2 V^{-1} s^{-1}$ with the band-like transport and can preserve the semiconducting properties, which can be used as photo-respond materials for photodetector applications. This work focuses on both synthesis and functionality. The new chemistry may shed light on organic semiconducting material preparation on the water surface

## Acknowledgments

This research was funded by the Slovenian Research And Innovation Agency, grant numbers P1-0055 and Z1-3189.